\documentclass{emulateapj}

\usepackage{graphicx}
\usepackage{txfonts}
\usepackage{longtable}
\usepackage{lscape}
\usepackage{natbib}

\begin{document}

\title{The physical origin and the diagnostic potential of the scattering 
polarization in\\ 
the lithium resonance doublet at 6708~{\AA}}

\author{{\normalsize Luca Belluzzi}\altaffilmark{1},
{\normalsize Egidio Landi Degl'Innocenti}\altaffilmark{2,1} 
{\normalsize and Javier Trujillo Bueno}\altaffilmark{2,3}}

\affil{$^1$Dipartimento di Astronomia e Scienza dello Spazio, 
Universit\`a di Firenze, Largo E. Fermi 2, I-50125 Firenze, Italy\\
$^2$Instituto de Astrof\'isica de Canarias, V\'ia L\'actea s/n, 
E-38205 La Laguna, Tenerife, Spain\\
$^3$Consejo Superior de Investigaciones Cient\'ificas, Spain}


\begin{abstract}
High-sensitivity measurements of the linearly-polarized solar limb spectrum 
produced by scattering processes in quiet regions of the solar atmosphere 
showed that the $Q/I$ profile of the lithium doublet at 6708~{\AA} has an 
amplitude ${\sim}10^{-4}$ and a curious three-peak structure, qualitatively 
similar to that found and confirmed by many observers in the Na~{\sc i} D$_2$ 
line. Given that a precise measurement of the scattering polarization profile 
of the lithium doublet lies at the limit of the present observational 
possibilities, it is worthwhile to clarify the physical origin of the observed 
polarization, its diagnostic potential and what kind of $Q/I$ shapes can be 
expected from theory. 
To this end, we have applied the quantum theory of atomic level 
polarization taking into account the hyperfine structure of the two stable 
isotopes of lithium, as well as the Hanle effect of a microturbulent  
magnetic field of arbitrary strength. We find that quantum interferences 
between the sublevels pertaining to the upper levels of the D$_2$ and D$_1$ 
line transitions of lithium do not cause any observable effect on the emergent 
$Q/I$ profile. Our theoretical calculations show that only two $Q/I$ peaks can 
be expected, with the strongest one caused by the D$_2$ line of $^7$Li~{\sc i} 
and the weakest one due to the D$_2$ line of $^6$Li~{\sc i}. 
Interestingly, we find that these two peaks in the theoretical $Q/I$ profile 
stand out clearly only when the kinetic temperature of the thin atmospheric 
region that produces the emergent spectral line radiation is lower than 4000 K.
The fact that such region is located around a height of 
200~km in standard semi-empirical models, where the kinetic temperature is 
about 5000~K, leads us to suggest that the most likely $Q/I$ profile produced 
by the sun in the lithium doublet should be slightly asymmetric and dominated 
by the $^7$Li~{\sc i} peak.
\end{abstract}


\section{Introduction}
\begin{figure}
\centering
\includegraphics[width=0.53\textwidth]{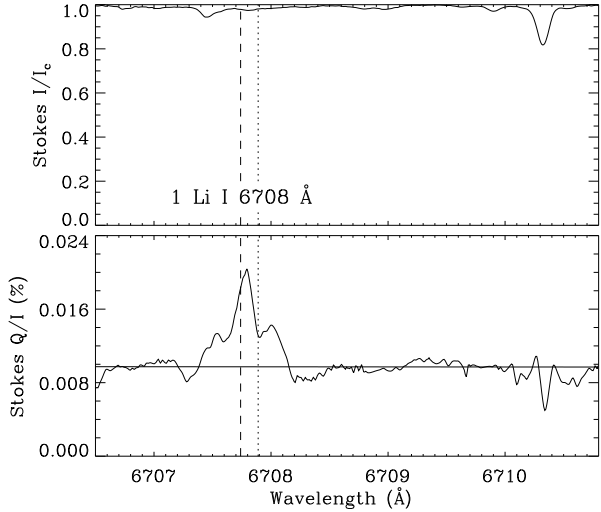}
\caption{Intensity spectrum (upper panel) and fractional linear polarization 
(lower panel) of the Li~{\sc i} 6708~{\AA} doublet, as observed by Stenflo et 
al. (2000). The wavelength positions of the 
two components of this doublet (which are of the D$_2$ and D$_1$ type), due to
the most abundant lithium isotope ($^7$Li), are indicated by the dashed and 
dotted lines, respectively. The recording was made on 1996 September 8, near 
the south polar limb. From \citet{Ste00}.}
\label{fig:observation}
\end{figure}

One of the interesting peculiarities of the linearly-polarized spectrum 
observed by \citet{Ste97} in quiet regions close to the solar limb is  
that a variety of spectral lines from minority species, which produce almost 
negligible absorption features in the Fraunhofer spectrum, nonetheless stand up 
with significant contrast in  fractional linear polarization (i.e., in the 
$Q(\lambda)/I(\lambda)$ spectrum)\footnote{Note that the first Stokes 
parameter, $I(\lambda)$, is the specific intensity at a given wavelength, while 
Stokes $Q(\lambda)$ represents here the intensity difference between linear 
polarization parallel and perpendicular to the closest solar limb.}. 
For example, molecules make a significant contribution to the structural 
richness of this so-called Second Solar Spectrum, and it is of interest to 
note that their magnetic sensitivity through the Hanle effect has facilitated 
the exploration of the sun's hidden magnetism \citep[see the detailed review 
by][and more references therein]{JTB06}. 
Spectral lines from rare earths, such as those from ionized cerium, are 
also of interest for obtaining information on unresolved, tangled magnetic 
fields in the ``quiet'' solar atmosphere \citep[see][]{Man06}. 
The main aim of this paper is to investigate the physical origin and the 
diagnostic potential of what is probably the weakest $Q/I$ spectral feature 
produced by a minority species in the whole Second Solar Spectrum: that 
observed by \citet{Ste00} in the Li~{\sc i} resonance doublet at 6708~{\AA} 
(see Figure~\ref{fig:observation}). 
Although it is generally accepted that the Second Solar Spectrum is due to 
radiatively induced population imbalances and quantum coherences in the atoms 
and molecules of the solar atmosphere, and several interesting spectral lines 
have been successfully modeled \citep[e.g.,][]{Man-JTB-03}, much remains to be 
done to fully understand the $Q/I$ profiles observed in many other spectral 
lines.

As seen in Figure~\ref{fig:observation}, the line center value of the observed 
$Q/I$ profile published by \citet{Ste00} is only $2{\times}10^{-4}$, 
and its detection required to sacrifice completely the spatio-temporal  
resolution in order to be able to push down the {\em rms} noise to below 
the $10^{-5}$ level. 
The FWHM of the observed $Q/I$ profile is about 190 m{\AA}. 
The most peculiar feature of the observed signal is its three-peak structure, 
which bears a qualitative resemblance to the triple peak structure of the $Q/I$ 
profile observed in the Na~{\sc i} D$_2$ line.  
Like the sodium doublet, the Li~{\sc i} one at 6708~{\AA} results from a 
D$_2$ type transition ($J_u=3/2$ and $J_{\ell}=1/2$) and a D$_1$ type  
transition ($J_u=1/2$ and $J_{\ell}=1/2$), with the difference that  
lithium has two isotopes: $^7$Li (with nuclear spin $I=3/2$ and a relative 
meteoritic abundance of 92.41\%) and $^6$Li (with $I=1$ and a relative 
meteoritic abundance of 7.59\%). 
The fact that quantum interferences between the $J_u=3/2$ and $J_u=1/2$ levels
are indeed important for understanding the $Q/I$ pattern observed at 
wavelengths around the sodium D$_2$ and D$_1$ lines \citep{Ste97,Lan98} led 
\citet{Ste00} to argue that the $Q/I$ feature shown in 
Figure~\ref{fig:observation} must be seriously influenced by a quantum 
mechanical superposition of the scattering transitions in the D$_2$ and D$_1$ 
lines of lithium.
\begin{table*}[t!]
\begin{center}
\caption{Physical properties of the two stable isotopes of lithium}
\label{tab:phys-prop}
\begin{tabular}{cccccccc}
\hline
\hline
\noalign{\smallskip}
Isotope & Abund. & $I$ & \multicolumn{2}{c}{Isotope Shifts (MHz)$^{a}$} & \multicolumn{3}{c}{HFS Constants (MHz)} \\
\noalign{\smallskip}
\cline{4-5}
\cline{6-8}
\noalign{\smallskip}
& (\%)$^{\rm{b}}$ & & D$_1$ & D$_2$ & $^2S_{1/2}$ & $^2P_{1/2}$ & $^2P_{3/2}$ \\
\noalign{\smallskip}
& & & & & $\mathcal{A}$ & $\mathcal{A}$ & $\mathcal{A}$ \\
& & & & & & & $\mathcal{B}$ \\
\hline
\noalign{\smallskip}
$^{6}$Li & 7.59 & 1 & -10533.13$^{\rm{c}}$ & -10534.93$^{\rm{c}}$ & 152.1368393$^{\rm{d}}$ & 17.375$^{\rm{e}}$ & -1.155$^{\rm{e}}$ \\
& & & & & & & -0.010$^{\rm{e}}$ \\
$^{7}$Li & 92.41 & 3/2 & \multicolumn{2}{c}{reference isotope} & 401.7520433$^{\rm{d}}$ & 45.914$^{\rm{f}}$ & -3.055$^{\rm{f}}$ \\
& & & & & & & -0.221$^{\rm{f}}$ \\
\noalign{\smallskip}
\hline
\end{tabular}
\end{center}
\small{$^{a}$A positive isotope shift means that the line is shifted to
higher frequencies with respect to the reference isotope.
$^{\rm{b}}$\citet{Ral08}; $^{\rm{c}}$\citet{Sch96};
$^{\rm{d}}$\citet{Bec74}; $^{\rm{e}}$\citet{Ort74}; $^{\rm{f}}$\citet{Ort75}.\\
\\}
\end{table*}

An accurate measurement of the very weak linear polarization signal of the  
Li~{\sc i} doublet at 6708~{\AA} is so difficult that we believe that it is 
very important to investigate carefully whether or not the $Q/I$ profile 
observed by \citet{Ste00} can be confirmed by theory.  
We think that this can be achieved with a particularly high level of 
confidence for this lithium doublet, because its extreme weakness in the 
intensity spectrum suggests that radiative transfer effects play no significant 
role on the {\em shape} of the $Q/I$ profile. 
Indeed, the anisotropic radiation pumping is basically that due to the solar 
continuum radiation. 
Moreover, since the incident radiation field is practically flat across the 
lithium doublet, we can rigorously apply the complete redistribution theory of 
spectral line polarization \citep[see the monograph by][hereafter referred to 
as LL04]{Lan04}, either neglecting or accounting for quantum interferences 
between the hyperfine structure (HFS) $F$-levels pertaining to the $J_u=3/2$ 
and $J_u=1/2$ levels.

As we shall see, the application of the above-mentioned quantum theory of 
spectral line polarization leads to the conclusion that only two peaks can be 
expected for the $Q/I$ profile of the Li~{\sc i} resonance doublet at 
6708~{\AA}. Interestingly, we find that the strongest peak at 6707.75~{\AA} 
is caused by the D$_2$ line of $^7$Li~{\sc i}, while the weakest one at about 
6707.9~{\AA} is due to the D$_2$ line of $^6$Li~{\sc i}.
Since in the sun the line opacity of this lithium doublet is negligible 
with respect to that of the continuum, contrary to the case of strong lines 
like Na~{\sc i} D$_1$ and D$_2$, or Ca~{\sc ii} H and K, these lithium lines 
neither show extended wings in the intensity spectrum nor in the $Q/I$
spectrum, the continuum level being rapidly reached as one moves away from  
line center. This explains why quantum interferences between HFS $F$-levels 
pertaining to different $J$-levels, whose signatures are negligible in the 
line core but become dominant in the wings, are found to play no significant 
role on the emergent $Q/I$ profile of this lithum doublet.
Moreover, we find that such two peaks in the theoretical $Q/I$ profile 
stand up clearly only when the kinetic temperature of the thin atmospheric 
region that produces the emergent $Q/I$ profile is lower than 4000~K. 
In order to be able to reproduce the width of the observed $Q/I$ profile 
without accounting for non-thermal broadening we need at least 6000~K, but for
T$\,{>}\,$5000 K the predicted $Q/I$ profile is slightly asymmetric (i.e., 
with a more extended red wing) and only shows the dominant peak due to 
$^7$Li~{\sc i}.

The outline of the paper is the following. In Section~2, we present the atomic 
model and the structure of the hyperfine multiplets, both of $^6$Li~{\sc i} 
and of $^7$Li~{\sc i}. 
In the same section, we briefly present the density matrix formalism which we
have adopted in order to describe the atomic polarization induced in the 
various levels by anisotropic radiation pumping processes.  
The optically thin slab model considered in this investigation, and the 
relevant equations describing the polarization of the emergent radiation 
are discussed in Section~3. In Section~4, the resulting theoretical $Q/I$ 
profile is shown, and the physical origin of its spectral features is 
investigated. 
Section 4.1 is dedicated to a discussion of the role of quantum 
interferences between the upper $J$-levels of the D$_1$ and D$_2$ lines.
Finally, the sensitivity of the resulting profile to the isotopic abundances  
and to microturbulent magnetic fields is discussed in Section~5. 
Section 6 collects our conclusions with an outlook to future research.

\section{The atomic model}
\begin{figure*}
\centering
\includegraphics[width=0.77\textwidth]{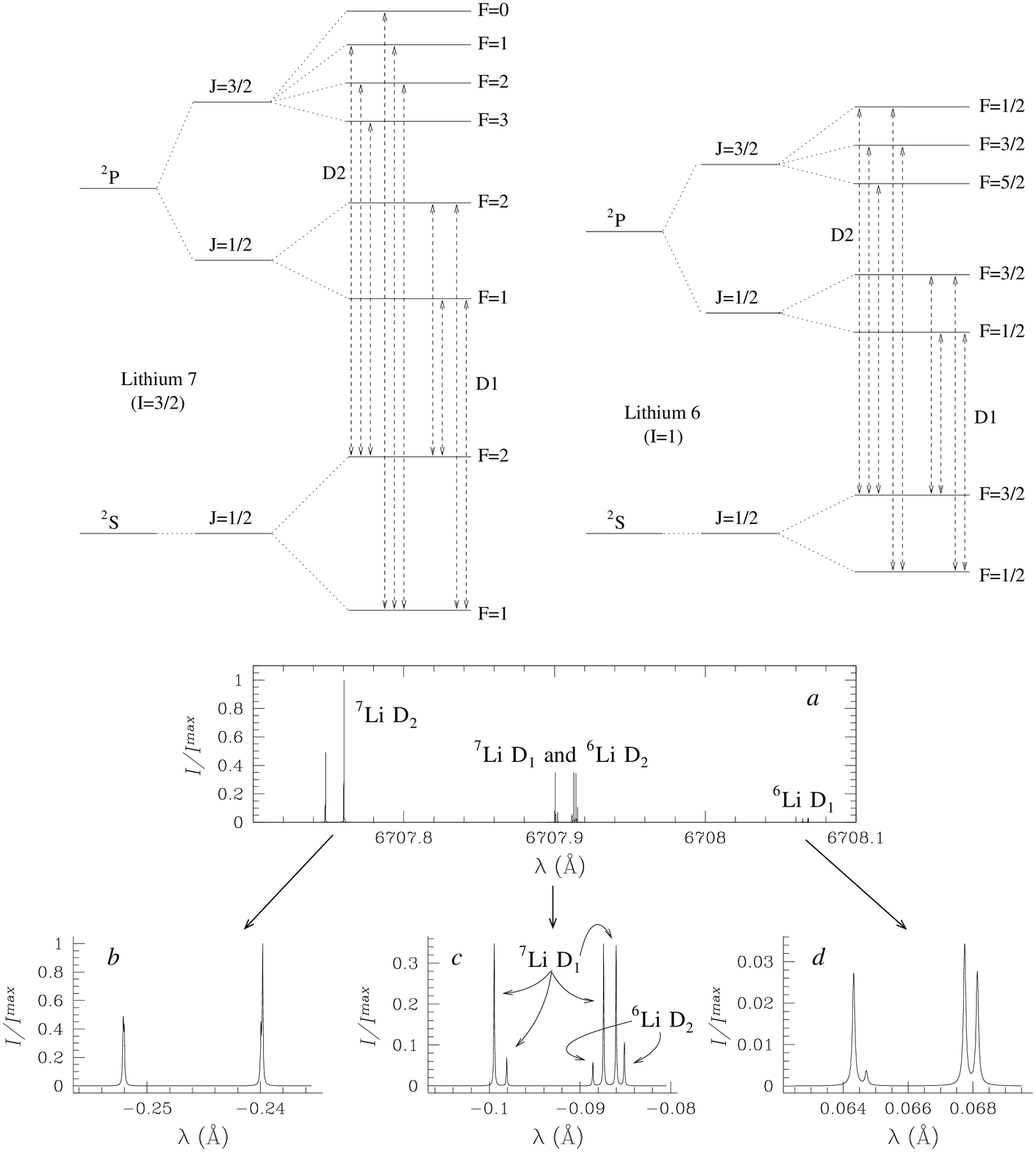}
\caption{Grotrian diagram showing the terms, the fine structure, and the 
hyperfine structure levels considered in our atomic models for the $^6$Li and 
$^7$Li isotopes (splittings are not drawn to scale). 
On the Grotrian diagrams the HFS components of the D$_1$ and 
D$_2$ lines of both isotopes are drawn. Panel {\it a} shows the laboratory 
positions and relative intensities of all the HFS components, as obtained 
taking into account the meteoritic relative abundance of the two isotopes. 
Panels {\it b}, {\it c} and {\it d} show in more detail the HFS components 
of the $^7$Li~{\sc i} D$_2$ line ({\it b}), the HFS components of the 
$^7$Li~{\sc i} D$_1$ line together with those of the $^6$Li~{\sc i} D$_2$ line 
({\it c}), and the HFS components of the $^6$Li~{\sc i} D$_1$ line ({\it d}). 
Note that the scale of the $I/I^{\rm max}$ graph is different in the various 
panels. The zero of the wavelength scale in panels {\it b}, {\it c} and {\it d} 
is taken at 6708~{\AA}.}
\label{fig:grotrian}
\end{figure*}
We adopt a three-level model of Li~{\sc i} consisting in the ground level 
(2{\it s} $^2$S$_{1/2}$), the upper level of the D$_1$ line 
(2$p$ $^2$P$_{1/2}$), and the upper level of the D$_2$ line 
(2$p$ $^2$P$_{3/2}$). The energies of the fine structure levels are taken 
from \citet{Ral08}. In order to take into account the isotopic effect, we 
correct the energies of the $^2$P$_{1/2}$ and of the $^2$P$_{3/2}$ levels of 
$^6$Li~{\sc i} using the values of the isotopic shifts in the D$_1$ and 
D$_2$ lines listed in Table~\ref{tab:phys-prop}.

The HFS Hamiltonian, describing the interaction between the nuclear spin and 
the electronic angular momentum, can be expressed as a series of electric and 
magnetic multipoles \citep[see, for example,][]{Kop58}. We calculate the 
energies of the HFS $F$-levels using the values of the magnetic dipole and 
of the electric quadrupole HFS constants (usually indicated with the symbols  
$\mathcal{A}$ and $\mathcal{B}$, respectively) listed in 
Table~\ref{tab:phys-prop}.
We recall that in the absence of magnetic fields, using Dirac's notation, the 
energy eigenvectors can be written in the form $|\alpha J I F f >$, where 
$\alpha$ represents a set of inner quantum numbers (specifying the 
configuration and, if the atomic system is described by the $L$-$S$ coupling
scheme, the total electronic orbital and spin angular momenta), $J$ is the 
total electronic angular momentum quantum number, while $F$ and $f$ are the 
quantum numbers associated with the total angular momentum operator 
(electronic plus nuclear: {\bf F}={\bf J}+{\bf I}), and with its projection
along the quantization axis, respectively.
The Grotrian diagrams showing the various HFS $F$-levels of the two isotopes, 
and the HFS components of the D$_1$ and D$_2$ lines are shown in the upper 
panel of Figure~\ref{fig:grotrian}. 
In the lower panels of Figure~\ref{fig:grotrian} the laboratory positions of 
the various HFS components are shown. Since the isotopic shifts are of the same 
order of magnitude as the frequency separation between the two D-lines, the 
D$_1$ line of $^7$Li~{\sc i} falls almost at the same wavelength as the 
D$_2$ line of $^6$Li~{\sc i} (see panels {\it a} and {\it c} of 
Figure~\ref{fig:grotrian}). 
We observe that in both isotopes the ground level splits into two HFS 
$F$-levels. The splitting between these two levels is much larger than that 
among the HFS $F$-levels pertaining to the upper levels. 
As a consequence, the HFS components both of the D$_1$ and of the D$_2$ lines 
can be gathered into two groups. Note also that it is not possible to resolve,
even in the laboratory spectra, all the HFS components of the D$_2$ line, since 
their frequency separation is smaller than the natural width (compare the 
magnetic dipole HFS constant of the D$_2$ upper level with the Einstein 
coefficient for spontaneous emission in this line, quoted in 
Table~\ref{tab:li-transitions}).

We describe the excitation state of the Li~{\sc i} levels by means of the 
density matrix formalism, a robust theoretical framework very suitable for 
treating the atomic polarization (population unbalances, and quantum 
interferences among the magnetic sublevels) that can be induced, for instance, 
by an anisotropic incident radiation field. 
In principle, to have a complete description of the atomic polarization, one 
has to take into account all the quantum interferences (or coherences) of the 
form
\begin{equation}
	< \alpha J I F f \, | \, \hat{\rho} \, | \, \alpha J^{\prime} I 
	F^{\prime} f^{\prime} > \;\; ,
\end{equation}
where $\hat{\rho}$ is the density operator.
In this investigation we, first, restricted to the $J$-diagonal density matrix 
elements
\begin{equation}
	< \alpha J I F f \, | \, \hat{\rho} \, | \, \alpha J I F^{\prime} 
	f^{\prime} > \;\; ,
\end{equation}
or, in other words, we neglected coherences between different $J$-levels. 
According to LL04, the resulting model atom is referred to as {\it multi-level 
atom with HFS}.
The statistical equilibrium equations (SEE), and the radiative transfer
coefficients for a multi-level atom with HFS can be found in LL04.
More general calculations, that we carried out including the interferences 
between the upper HFS levels of the D$_1$ and D$_2$ lines, showed that
the previous approximation is fully justified for the investigation of this 
lithium doublet.
Indeed, as it will be shown in Section~4.1, the theoretical $Q/I$ profiles 
obtained through the solution of these more general equations cannot be 
distinguished from those shown in this paper, which, as mentioned above, 
neglect quantum interferences between different $J$-levels.
This is due to the weakness of the lithium doublet which does not 
allow for the formation of extended wings where such quantum interferences 
would produce their main signatures.

\section{The optically thin slab model}
\begin{table}[!t]
\centering
\caption{Wavelengths (in air) and Einstein coefficients of the lines 
considered; mean number of photons and anisotropy factor of the photospheric 
continuum at the wavelengths of the same lines.}
\smallskip
\begin{tabular}{ccccc}
\hline
\hline
\noalign{\smallskip}
Line & $\lambda$~(\AA) & $A\,(s^{-1})$ & $\bar{n}_{\nu}$ & $w_{\nu}$ \\
\noalign{\smallskip}
\hline
\noalign{\smallskip}
 D$_1$ & 6707.91 & 3.72$\times 10^7$ & 0.0115 & 0.0992 \\
 D$_2$ & 6707.76 & 3.72$\times 10^7$ & 0.0115 & 0.0992 \\
\noalign{\smallskip}
\hline
\noalign{\smallskip}
\end{tabular}
\label{tab:li-transitions}
\end{table}
In order to emphasize the atomic aspects involved in the problem, avoiding
complications due to radiative transfer effects, we consider a horizontal, 
optically thin slab of Li~{\sc i} ions. 
We assume the slab to be illuminated from below by the observed photospheric continuum
radiation field, that we suppose to be unpolarized and cylindrically symmetric
about the local vertical. Under these assumptions, taking a reference system 
with the $z$-axis (the quantization axis) directed along the vertical, only 
two components of the radiation field tensor, through which we describe the 
incident continuum radiation, are non-vanishing:
\begin{equation}
J^{0}_{0}(\nu)\! \! =\! \! \oint\frac{{\rm d}\Omega}{4\pi}I(\nu,\mu)
\;\;\; {\rm and} \;\;\;\;
J^{2}_{0}(\nu)\! \! =\! \! \oint\frac{{\rm d}\Omega}{4\pi}\frac{1}
{2\sqrt{2}} (3\mu^2-1)I(\nu,\mu) \;\; ,
\label{eq:JKQ}
\end{equation}
where $\mu$ is the cosine of the heliocentric angle. 
The former quantity is the mean intensity of the incident radiation field 
(averaged over all directions), the second one quantifies its degree of 
anisotropy (unbalance between vertical and horizontal illumination).
Mean intensity and anisotropy degree of the radiation field can also be 
expressed in terms of two non-dimensional quantities: the average number 
of photons per mode, $\bar{n}$, and the so-called anisotropy factor, $w$ 
(which varies between $w=-1/2$, for the case of horizontal illumination by an 
unpolarized radiation field without azimuthal dependence, and $w=1$ for 
vertical unpolarized illumination). 
Such quantities are related to $J^0_0$ 
and $J^2_0$ by the equations
\begin{equation}
\bar{n}(\nu)=\frac{c^2}{2 h \nu^3} J^0_0(\nu) \;\; , \;\;\;
w(\nu)=\sqrt{2} \frac{J^2_0(\nu)}{J^0_0(\nu)} \;\; .
\end{equation}
We calculate the values of $\bar{n}(\nu)$ and $w(\nu)$ of the photospheric
continuum at the frequencies of the Li~{\sc i} D-lines, following Section~12.3 
of LL04 taking $h=0$, and using the values of the disk-center 
intensities and of the limb-darkening coefficients given by \citet{Pie00}. 
The values obtained are listed in Table~\ref{tab:li-transitions}.
Once the SEE have been written down and solved numerically, we can calculate 
the radiative transfer coefficients. 
We consider the radiation scattered by the slab at 90$^{\circ}$, and we take 
the reference direction for positive $Q$ parallel to the slab.
For the case of a tangential observation, in a weakly polarizing atmosphere
($\varepsilon_I \gg \varepsilon_Q, \varepsilon_U,\varepsilon_V; \eta_I \gg 
\eta_Q, \eta_U, \eta_V, \rho_Q, \rho_U, \rho_V$), the polarization of the 
emergent radiation is given by \citep[see][]{JTB03}
\begin{equation}
	\label{eq:dichroism}
	\frac{X(\nu,\mathbf{\Omega})}{I(\nu,\mathbf{\Omega})} \approx
	\frac{\varepsilon_{X}(\nu,\mathbf{\Omega})}
	{\varepsilon_{I}(\nu,\mathbf{\Omega})}-
	\frac{\eta_{X}(\nu,\mathbf{\Omega})}
	{\eta_{I}(\nu,\mathbf{\Omega})} \;\;\;\;\;\;{\rm{with}}\; X=Q,U,V \;\; .
\end{equation}
The first term in the right hand side of Eq.~(\ref{eq:dichroism}) represents
the contribution to the emergent radiation due to selective emission processes,
the second one is caused by dichroism (selective absorption of polarization
components).

The way atomic polarization is distributed among the various levels by 
radiative processes is very similar to the case of the sodium and barium 
D-lines, investigated by \citet{JTB02} and by \citet{Bel07}, respectively.
We recall in particular that only the upper level of the D$_2$ line can be 
directly polarized by the anisotropic incident radiation field. The ground 
level becomes polarized because of a transfer of atomic polarization via 
spontaneous emission in the D$_2$ line, while the D$_1$ upper level is 
polarized via radiative absorption in the D$_1$ line (``repopulation 
pumping'').
Indeed, solving the SEE in the absence of magnetic fields, one finds that 
the D$_2$ line upper level is considerably more polarized than both the 
D$_1$-line upper level and the ground level.
Dichroism therefore can be safely neglected as far as D$_2$ is concerned, 
while it is expected to be more important in the D$_1$ line. However, given 
that the polarization signal produced by the D$_1$ line of Li~{\sc i} is 
expected to be several orders of magnitude smaller than the one produced by 
the D$_2$ line, we neglect dichroism also in the D$_1$ line. 
We calculate therefore the polarization of the emergent radiation through the 
simpler equation
\begin{equation}
	\label{eq:QsuI}
	\frac{X(\nu,\mathbf{\Omega})}{I(\nu,\mathbf{\Omega})} \approx
	\frac{\varepsilon_{X}(\nu,\mathbf{\Omega})}
	{\varepsilon_{I}(\nu,\mathbf{\Omega})} \;\; .
\end{equation}
The emission coefficients that appear in the previous equation include only 
line processes: in order to reproduce the observed $Q/I$ profile, we need to 
add the contribution of the continuum. Assuming such contributions to be constant across 
the line we have
\begin{equation}
	\label{eq:QsuI_cont}
	\frac{X(\nu,\mathbf{\Omega})}{I(\nu,\mathbf{\Omega})} \approx
	\frac{\varepsilon_{X}^{\,\ell}(\nu,\mathbf{\Omega})+\varepsilon_X^{\,c}}
	{\varepsilon_{I}^{\,\ell}(\nu,\mathbf{\Omega})+\varepsilon_{I}^{\,c}} 
	\;\; ,
\end{equation}
where the superscripts ``$c$'' and ``$\ell$'' recall that the corresponding
quantities refer to continuum and line processes, respectively.
The quantities $\varepsilon_{X}^{\, \rm c}$ and $\varepsilon_I^{\,\rm c}$ will 
be considered as free parameters to be adjusted in order to reproduce the 
observed polarization profile.
The effect of collisions will be neglected throughout this investigation.
It should be observed that given the extreme weakness of the absorption 
features of this lithium doublet in the solar intensity spectrum, the optically 
thin slab model is expected to be a rather good approximation for the modelling 
of these lines.

\section{Theoretical linear polarization profile of the emergent radiation}
\begin{figure}
\centering
\includegraphics[width=0.4\textwidth]{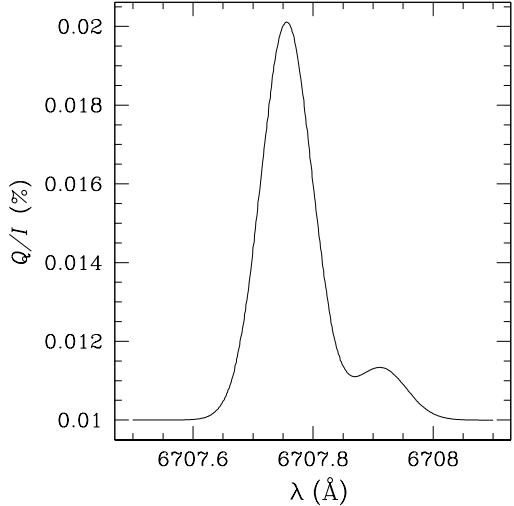}
\caption{Theoretical $Q/I$ profile obtained assuming the following values of 
the free parameters: 
$\varepsilon_I^{\, \rm c}=200 \times \varepsilon_I^{\, \rm max}$,
$\varepsilon_Q^{\, \rm c}= 10^{-4} \times \varepsilon_I^{\, \rm c}$. 
The profile is obtained for a Doppler width of 60~m{\AA}, which corrresponds 
to $T\approx3000$~K.}
\label{fig:best_fit}
\end{figure}

\begin{figure*}
\centering
\includegraphics[width=0.8\textwidth]{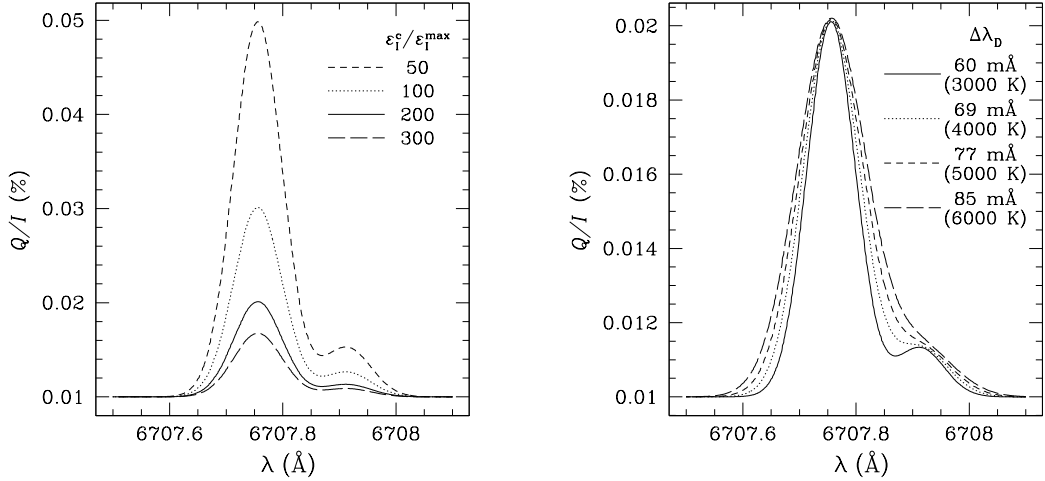}
\caption{Left panel: theoretical $Q/I$ profiles obtained for different values 
of the continuum ($\varepsilon_I^{\, \rm c}$). 
The solid-line profile is the same as in Figure~\ref{fig:best_fit}.
Right panel: theoretical $Q/I$ profiles obtained for different values of the 
Doppler width ($\Delta \lambda_D$).
The solid-line profile is the same as in Figure~\ref{fig:best_fit}.}
\label{fig:cont-doppler}
\end{figure*}

Applying Eq.~(\ref{eq:QsuI_cont}), and assuming the following values of the 
free parameters 
$\varepsilon_I^{\, \rm c}=200 \times \varepsilon_I^{\, \rm max}$, 
where $\varepsilon_I^{\, \rm max}$ is the maximum value of 
$\varepsilon_I^{\, \ell}$ 
in the wavelength range considered, $\varepsilon_Q^{\, \rm c}=10^{-4} \times 
\varepsilon_I^{\, \rm c}$, and a Doppler width of 60 m\AA, 
we obtain the two-peak linear polarization profile 
shown in Figure~\ref{fig:best_fit}. 
As can be observed in the left panel of Figure~\ref{fig:cont-doppler}, 
modifying the continuum contribution to the intensity we modify the amplitude 
of the two peaks, without changing the shape of the profile. 

The value that we assumed for the continuum contribution 
($\varepsilon_I^{\, \rm c}=200 \times \varepsilon_I^{\, \rm max}$) appears to 
be a good choice, given the weakness of 
the line, and that it allows the peak falling at shorter wavelengths (hereafter 
referred to as the ``blue peak'') to reach the same amplitude (0.02\%) as the 
central peak of the $Q/I$ profile observed by \citet{Ste00} (see 
Figure~\ref{fig:observation}). 
The value of $\varepsilon_Q^{\, \rm c}$ has been adjusted in 
order to obtain in the far wings the same continuum polarization level as in 
the observation (0.01\%).

Particularly interesting is the sensitivity of the theoretical $Q/I$ profile 
to the value of the Doppler width. The profile shown in 
Figure~\ref{fig:best_fit} has been obtained assuming a Doppler width of 
60~m{\AA}, which corresponds, neglecting microturbulent velocities, to a 
temperature of 3000~K. 
For the sake of simplicity, we consider the same Doppler width for the two 
isotopes, despite their mass difference ($\approx 14\%$)\footnote{Note that 
calculations made taking into account the mass difference between the two 
isotopes have not shown any appreciable modification of the results.}.
Profiles obtained assuming different values of the Doppler width are plotted in 
the right panel of Figure~\ref{fig:cont-doppler}. 
Interestingly, we observe that the two-peak structure gradually disappears as 
the Doppler width is increased: the observation of a two-peak structure 
could thus provide precise information concerning the thermal properties of 
the thin atmospheric region where this weak lithium doublet is formed.

\begin{figure*}
\centering
\includegraphics[width=0.8\textwidth]{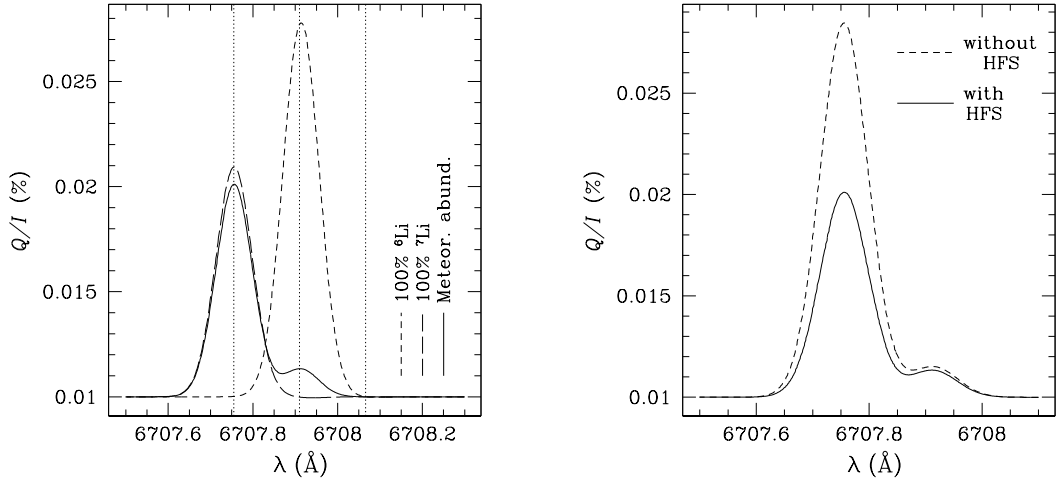}
\caption{Left panel: theoretical $Q/I$ profiles obtained ({\it a}) considering 
both isotopes weighted by their meteoritic abundances (solid line), 
({\it b}) assuming that only $^6$Li (100\% in abundance) is present 
(short-dashed line), ({\it c}) assuming that only $^7$Li (100\% in abundance) 
is present (long-dashed line). 
The vertical dotted lines show the wavelength positions of the HFS components 
of the $^7$Li~{\sc i} D$_2$ line (first line going from shorter to longer 
wavelengths), of the $^7$Li~{\sc i} D$_1$ line together with the $^6$Li~{\sc i} 
D$_2$ line (middle line), and of the $^6$Li~{\sc i} D$_2$ line (last line). 
The solid-line profile is the same as in Figure~\ref{fig:best_fit}.
Right panel: theoretical $Q/I$ profiles obtained considering both isotopes, 
weighted by their meteoritic abundances, taking into account (solid line) 
and neglecting (dashed line) HFS.
The solid-line profile is the same as in Figure~\ref{fig:best_fit}.
All profiles have been obtained for a Doppler width of 60~m{\AA}.
}
\label{fig:2is-hfs}
\end{figure*}

In the left panel of Figure~\ref{fig:2is-hfs}, the same $Q/I$ profile as in 
Figure~\ref{fig:best_fit} is plotted together with the profiles 
expected in the hypothetical cases where only $^6$Li (short-dashed line) or 
only $^7$Li (long-dashed line) were present. The polarization peak that is 
obtained when a single isotope is considered (either $^6$Li or $^7$Li) is, in 
both cases, due to the corresponding D$_2$ line. 
Indeed, the polarization signals produced by the D$_1$ lines, both of 
$^6$Li~{\sc i} and of $^7$Li~{\sc i}, are several orders of magnitude smaller, 
and cannot be appreciated on this plot. 
It is then clear that the physical origin of the two-peak structure of 
the $Q/I$ profile that we have obtained within our modelling assumption lies in 
the isotopic shift between the two lithium isotopes: the two peaks are nothing 
else but the signals produced by the D$_2$ lines of $^7$Li~{\sc i} (blue peak) 
and of $^6$Li~{\sc i} (red peak), which fall at different wavelengths because 
of the isotopic shift, and which are weighted by the isotopic abundances.

It is also interesting to observe that the signals due to the D$_2$ lines 
of the two isotopes, as found in the hypothetical cases where only $^6$Li or 
only $^7$Li were present, do not have the same amplitude (see the left panel of 
Figure~\ref{fig:2is-hfs}).
This is due to the fact the two isotopes have different HFS (in particular 
different nuclear spin quantum numbers, and different HFS constants, producing 
different splittings among the various $F$-levels, so that $^6$Li is less 
depolarized by HFS than $^7$Li). 
As a consequence, the relative amplitude of the two peaks that we have found 
in the resulting $Q/I$ profile does not reflect only the different abundances 
of the two lithium isotopes, but also their different HFS. 
Therefore, the presence of HFS in the two isotopes has to be taken into account 
in order to obtain the correct relative amplitude between the two peaks. 
This is clearly shown in the right panel of Figure~\ref{fig:2is-hfs}, where the 
profiles obtained including (solid line) and neglecting (dashed line) HFS are 
plotted.

\subsection{The role of coherences between the upper $J$-levels of the 
Li~{\sc i} D$_1$ and D$_2$ lines}
\begin{figure}
\centering
\includegraphics[width=0.5\textwidth]{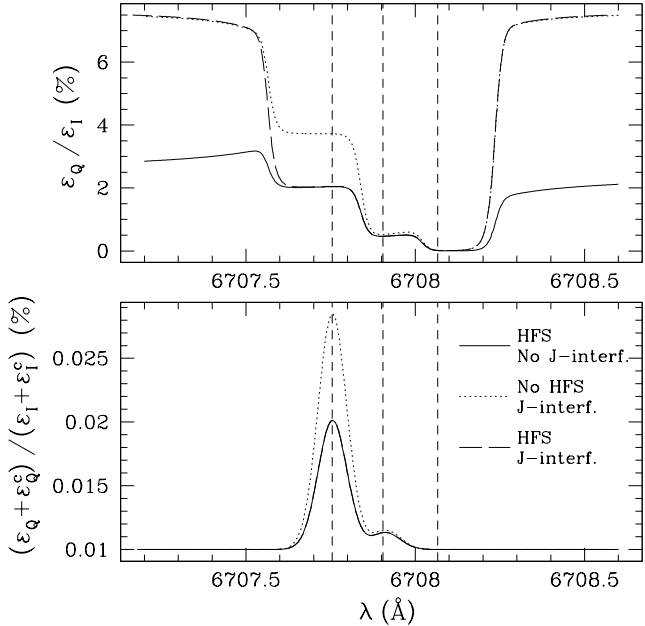}
\caption{Upper panel: theoretical fractional polarization profiles obtained in 
the absence of continuum according to three different approximations (see the
text). The quantities $\bar{n}$, $w$, and $\Delta \lambda_D$ have the same 
values as in Figure~\ref{fig:best_fit}. The vertical dashed lines show the 
wavelength positions (going from the blue to the red) of $^7$Li~{\sc i} D$_2$, 
of $^7$Li~{\sc i} D$_1$ and $^6$Li~{\sc i} D$_2$ (blended), and of 
$^6$Li~{\sc i} D$_1$. 
Lower panel: same as the upper panel, but including the contribution of the 
continuum. The quantities $\varepsilon_I^{\, \rm c}$ and 
$\varepsilon_Q^{\, \rm c}$ have the same values as in 
Figure~\ref{fig:best_fit}. The profile plotted with solid line is the same as 
in Figure~\ref{fig:best_fit}.}
\label{fig:super-interf}
\end{figure}

Although the D$_1$ and D$_2$ lines (both of $^6$Li~{\sc i} and $^7$Li~{\sc i}) 
are very close to each other, so that quantum interferences between the upper 
levels of the corresponding transitions are not negligible, it is important to 
point out that, because of the weakness of these lines, the continuum level is 
rapidly reached both in the intensity spectrum, and in the $Q/I$ spectrum, and 
the spectral signatures of such coherences are completely lost.
This can be easily appreciated in Figure~\ref{fig:super-interf}, where the 
theoretical $Q/I$ profiles, obtained both without (upper panel) and with (lower 
panel) continuum, are plotted according to three different approximations: 
{\it a}) taking into account HFS, but neglecting interferences between pairs
of HFS magnetic sublevels pertaining to different $J$-levels (solid line), 
{\it b}) neglecting HFS, but taking into account interferences between 
pairs of magnetic sublevels pertaining to different $J$-levels (dotted line), 
and  {\it c}) taking into account HFS, and taking also into account 
interferences between pairs of HFS magnetic sublevels pertaining to different 
$J$-levels (dashed line).
The first approximation is the one that has been applied in this paper
(see Section~2). A detailed description of the other approximations, usually 
referred to as {\it multi-term atom} and {\it multi-term atom with HFS} can be 
found in LL04, and in \citet{Cas05}, respectively.

We start considering the results obtained without continuum (upper panel).
As expected, close to line center there is no appreciable difference between 
the profile obtained including the interferences between the upper levels of 
the D$_1$ and D$_2$ lines (dashed-line profile) and the one obtained neglecting 
such interferences (solid-line profile). 
Actually, we recall that such interferences play an important role in the 
wings of the line, being negligible close to the line core.
At line center, on the other hand, there is an appreciable difference 
between the profiles obtained taking into account (solid- and dashed-line 
profiles) and neglecting (dotted-line profile) HFS. 
In the far wings, on the contrary, as can be observed by comparing the dashed- 
and the dotted-line profiles, there is no difference between the results 
obtained by taking into account or neglecting HFS (as known, the effect of HFS 
vanishes in the far wings of the line)\footnote{Note that this is strictly true 
only if the lower level is not polarized, or if it carries an amount of atomic 
polarization much smaller than that of the upper level, as in the lines under 
investigation.}. 
At these wavelengths, on the other hand, the interferences between different 
$J$-levels become important, and their effect can be easily appreciated. 
In particular, we observe that the asymptotic value reached by the profiles 
obtained taking into account such interferences (dashed- and dotted-line 
profiles) is much higher than the one reached by the profile calculated 
neglecting these interferences (solid-line profile).
However, if the continuum contributions to Stokes $I$ and $Q$ are included, so 
that a two-peak profile with the same amplitude as the observed one is 
obtained, the polarization in the wings of the line is rapidly decreased to the
continuum level, so that the profile obtained by taking into account the 
interferences between different $J$-levels cannot be distinguished from the 
one obtained neglecting such interferences (see the lower panel of 
Figure~\ref{fig:super-interf}, where the solid and dashed lines superimpose).
The only difference that can still be noticed is at line center, between the 
profiles obtained by including and neglecting HFS (as already shown in the 
left panel of Figure~\ref{fig:2is-hfs}).
We point out that the depolarization of the $Q/I$ profile in the wings 
of the line, which has completely hidden the signatures of the interferences 
between the different $J$-levels, is particularly strong in these lines since 
a significant continuum contribution 
($\varepsilon_I^{\, \rm c}=200 \times \varepsilon_I^{\, \rm max}$) is needed in 
order to reproduce the amplitude ($\approx 10^{-4}$) of the observed $Q/I$ 
profile.
Nevertheless, as already pointed out, this is a rather realistic value, given 
the extreme weakness of this lithium doublet\footnote{Note that also for 
smaller values of $\varepsilon_I^{\rm \, c}$ the effects of the interferences 
between different $J$-levels turn out to be hidden.}.
In conclusion, as anticipated in Section~2, interferences between the upper 
levels of the $D_1$ and D$_2$ lines, though present, can be safely neglected 
in the modelling of these lines.

\section{Sensitivity to the isotopic abundance and to a microturbulent magnetic 
field}
\begin{figure*}
\centering
\includegraphics[width=0.8\textwidth]{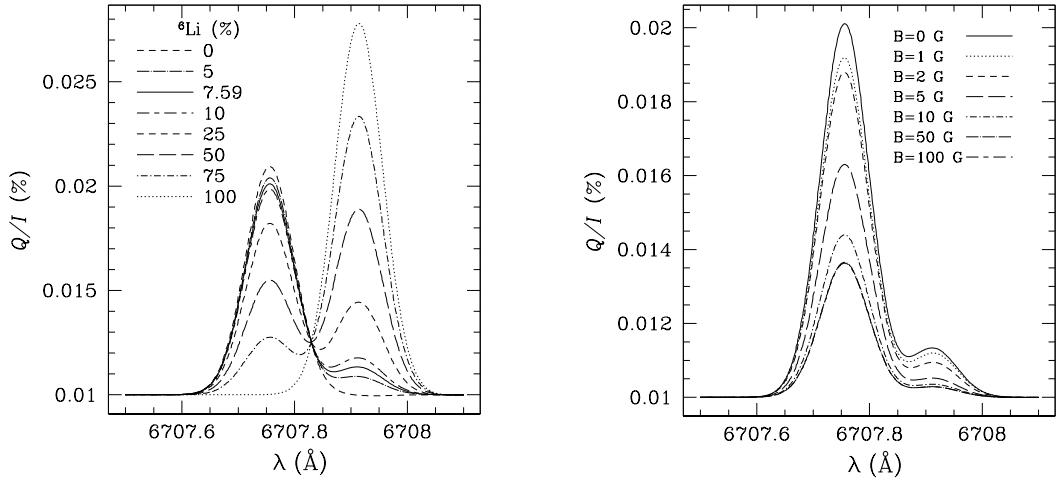}
\caption{Left panel: theoretical $Q/I$ profiles obtained for different isotopic
abundances. Right panel: theoretical $Q/I$ profiles obtained in the presence
of a microturbulent magnetic field of various intensities, assuming the
meteoritic abundances.
All the profiles are obtained through Eq.~(\ref{eq:QsuI_cont}), taking the 
values of the parameters specified in Section~4. 
The solid-line profile, both in the left and in the right panel, is the same 
as in Figure~\ref{fig:best_fit}.
All profiles have been obtained for a Doppler width of 60~m{\AA}.
}
\label{fig:abb-mag}
\end{figure*}

In the left panel of Figure~\ref{fig:abb-mag} the profiles obtained changing 
the relative abundance of the two isotopes are plotted.
We observe that as the $^6$Li abundance is increased, the amplitude of the
blue peak decreases, while the amplitude of the red one increases.
This behavior suggests the possibility of using the relative amplitude of the 
two peaks to estimate the abundance ratio of the two lithium isotopes in the 
quiet solar atmosphere.
Note that the two peaks have a slightly different sensitivity to the
relative isotopic abundance.
This is probably due to the fact that the blue peak is only caused by the
$^7$Li~{\sc i} D$_2$ line, whereas the red peak, though dominated by the 
$^6$Li~{\sc i} D$_2$ line, is also affected by the $^7$Li~{\sc i} D$_1$ line.
Obviously the amplitude of the two peaks of this $Q/I$ profile is not sensitive
only to the isotopic abundance, but also to collisions (here neglected),
and to the presence of a magnetic field.
As it can be observed in the right panel of Figure~\ref{fig:abb-mag}, in the
presence of a unimodal microturbulent and isotropic magnetic field the amplitude of the two 
peaks is reduced because of the Hanle effect.
Note that the two peaks have a slightly different magnetic sensitivity
(for example, the relative depolarization that takes place going from 1 to 2~G
is larger in the red peak). This is due to the fact that
the two isotopes have different nuclear spin quantum numbers and different
HFS splittings.
The saturation regime is reached for magnetic fields of about 50~G.
In conclusion, a correct estimate of the lithium isotopic abundance from the
relative amplitude of the two peaks of the $Q/I$ profile would require
an independent determination of the magnetic field strength present in the
regions of the solar atmosphere that produce the observed scattering
polarization. Likewise, for a given approximate value of the lithium isotopic
abundance the two peaks of the $Q/I$ profile of the lithium doublet can be
used to obtain information on the strength of unresolved, hidden magnetic
fields in the ``quiet'' sun. The problem is that it is unlikely that such two 
peaks can indeed be observed (see below).

\section{Concluding comments}

The application of the quantum theory of spectral line polarization leads to 
the conclusion that only two peaks can be expected for the $Q/I$ profile of 
the Li~{\sc i} resonance doublet at 6708~{\AA}. 
Interestingly, we find that the strongest peak at about 6707.75~{\AA} is caused
by the D$_2$ line of $^7$Li~{\sc i}, while the weakest one at about 
6707.9~{\AA} is due to the D$_2$ line of $^6$Li~{\sc i}. 
Since for the lithium doublet in the sun the line opacity is negligible with 
respect to that of the continuum, we find that quantum interferences between 
the HFS levels pertaining to different $J$-levels play no significant role on 
the emergent $Q/I$ profile. 
Moreover, we have showed that such two peaks in the theoretical $Q/I$ profile 
stand up clearly only when the kinetic temperature of the thin atmospheric 
region that produces the main contribution to the emergent $Q/I$ profile is 
sufficiently low (e.g., sensibly lower than 4000~K). 
In order to be able to reproduce the width of the observed $Q/I$ profile, 
without accounting for non-thermal broadening, we need at least 6000~K, but for 
this kinetic temperature the predicted $Q/I$ profile is slightly asymmetric 
(i.e., with a more extended red wing) and shows only the dominant peak due to 
$^7$Li~{\sc i}.
In this respect, it is of interest to note that the height in standard 
semi-empirical models of the solar atmosphere where the continuum optical 
depth at the wavelength of the lithium doublet is unity for a line-of-sight 
with $\mu=0.1$ is slightly below 200 km \citep[see Figure~6 of][]{JTB09}, where 
the kinetic temperature is about 5000 K. 
For this reason, we believe that the most likely {\em shape} of the scattering 
polarization profile of the lithium doublet that the sun can produce should be 
similar to the short-dashed line of the right panel of 
Figure~\ref{fig:cont-doppler}. 

If that is the case (i.e., if the true $Q/I$ profile of the lithium doublet 
turns out to have only one peak) then it would not be possible to apply the 
``peak-ratio technique'' illustrated in Figure~\ref{fig:abb-mag}, neither for 
determining the relative lithium isotopic abundances nor for estimating the 
strength of an unresolved, hidden magnetic field in the solar photosphere. 
Nevertheless, the amplitude, FWHM, and asymmetry of the observed $Q/I$ profile 
could still be used for such a purpose through radiative transfer simulations 
in a realistic model of the thermal and density structure of the quiet solar 
photosphere. 
For example, if the relative lithium isotopic abundance is known beforehand, 
the discrepancy between the $Q/I$ profile calculated in a three-dimensional 
hydrodynamical model of the solar photosphere and the observed profile could 
be interpreted in terms of an unresolved magnetic field, assuming that the 
depolarizing impact of elastic collisions is properly taken into account in 
the calculations. 
Since the relevant pumping radiation field here is that of the continuum 
radiation, whose anisotropy factor is significantly larger above the granule 
cell centers than above the intergranular lanes 
\citep[see Figure~2 of][]{JTB04}, the observed $Q/I$ profile of the lithium 
doublet (which lacks spatio-temporal resolution) must be significantly biased 
towards the granule cell centers. 
Therefore, we anticipate that the inferred magnetic field strength would be 
more representative of the magnetization of the granular plasma. 

Obviously, new and very careful observations of this extremely weak 
polarization signal are urgently needed. 
In this respect, it is of interest to mention that our colleagues from the 
Istituto Ricerche Solari Locarno (IRSOL), Dr. M. Bianda and Dr. R. Ramelli, 
have already initiated such an observational program with the 
Z\"urich Imaging Polarimeter (ZIMPOL) attached to the Gregory Coud\'e 
Telescope of IRSOL.
Although some of the $Q/I$ profiles they have been able to measure with such 
an excellent facility for high-sensitivity spectropolarimetry seem to 
show a one-peak profile, many more observations are needed in order to fully 
clarify what type of $Q/I$ profiles the sun is actually producing in the 
lithium doublet and in order to use them for diagnostic purposes.

\acknowledgements
The authors wish to thank Dr. Roberto Casini for providing a code that has 
been very useful for the computations contained in Section~4.1.
One of us, ELD, wishes to express his gratitude to the
Instituto de Astrof\'\i sica de Canarias (IAC) for finantial support 
that facilitated a sabatical stay at the IAC  
during 2009. Financial support by the Spanish Ministry of
Science through project AYA2007-63881 is also gratefully
acknowledged.

\end{document}